\documentclass[preprint,12pt]{elsarticle}

\usepackage{amsmath,amssymb}
\usepackage{booktabs}
\usepackage{tabularx}
\usepackage{array}
\usepackage{adjustbox}
\usepackage{xcolor}
\usepackage{microtype}
\usepackage{placeins}
\usepackage{tikz}
\usepackage{pgfplots}
\usetikzlibrary{arrows.meta,positioning,shapes.geometric,fit}
\usepgfplotslibrary{groupplots}
\pgfplotsset{compat=1.18}

\definecolor{dtcolor}{RGB}{31,119,180}
\definecolor{rfcolor}{RGB}{180,40,40}
\definecolor{xgbcolor}{RGB}{45,135,70}

\journal{Internet of Things}

\begin{document}

\begin{frontmatter}

\title{Beyond Detection Accuracy: Measuring Explanation Cost, Stability, and Utility for Resource-Aware IoT Intrusion Detection}

\author[aff1]{Abdurrahman Tolay\corref{cor1}}
\ead{abdurrahman.tolay@outlook.com}
\affiliation[aff1]{%
  organization={Independent Researcher},
  city={Istanbul},
  country={Türkiye}
}
\cortext[cor1]{Corresponding author}

\begin{abstract}
Machine-learning intrusion-detection studies commonly emphasize predictive accuracy while treating explanation generation as a computationally free post-processing step. This study jointly evaluates predictive effectiveness, explanation cost, local explanation stability, and selective explanation for binary Internet of Things (IoT) intrusion detection. A leakage-safe CICIoT2023 corpus was constructed with respect to exact 39-feature hashes through non-finite-value handling, exact-feature deduplication, conservative original-label collision removal, and deterministic hash-level partitioning. Logistic Regression, Decision Tree, Random Forest, and XGBoost were evaluated on natural and balanced test distributions. The computational cost of tree-based Shapley additive explanations (TreeSHAP) was measured, stability was assessed under prediction-preserving perturbations, and validation-calibrated policies were used to allocate explanation workload. XGBoost provided the strongest overall predictive profile, while Random Forest produced the lowest false-positive rate. Explanation cost differed sharply by architecture: at 5,000 samples, TreeSHAP required 700.759~s for Random Forest and 1.471~s for XGBoost. Random Forest showed the strongest overall base-level explanation stability, while Decision Tree also remained highly stable in top-feature membership and attribution rank; XGBoost retained high rank and directional consistency but exhibited greater top-feature turnover and attribution-magnitude drift. On the balanced test, approximately 90\% false-negative explanation coverage permitted compute savings of 28--32\%, while approximately 95\% coverage permitted savings of 15--23\%. Savings were substantially smaller under the attack-heavy natural prevalence. These results show that the operational value of explainable IoT intrusion detection depends on predictive quality, architecture-dependent explanation cost, local stability, workload prevalence, and selective invocation rather than on detection accuracy or explanation availability alone.
\end{abstract}


\end{frontmatter}

\section{Introduction}

Machine-learning intrusion-detection systems (IDSs) for Internet of Things (IoT) networks are commonly compared through predictive metrics, even though validation strategy, traffic composition, and deployment location materially affect the interpretation of those measurements \cite{khraisat2021iotids,moustafa2023explainable}. This evaluation is necessary, but it is not sufficient for selecting a model for deployment at an edge gateway or another resource-constrained monitoring point. An operational detector must identify attacks while limiting false alarms, meet latency and memory constraints, and provide information that supports investigation when a decision requires human review. These requirements concern different properties of the system and cannot be inferred from accuracy alone.

The distinction is particularly important for highly imbalanced intrusion-detection data. When attack observations dominate a test distribution, a model can attain high accuracy while still misclassifying a substantial proportion of benign traffic. Such false positives consume analyst attention and may make an otherwise accurate detector impractical. Precision, recall, F1 score, false-positive rate, and precision--recall area under the curve (PR-AUC) therefore provide more informative views of security performance, provided that they are interpreted in relation to the class distribution on which they were measured.

Post-hoc explanation introduces a second deployment issue. Methods such as LIME and SHAP are designed to characterize individual predictions, and TreeSHAP specializes additive attribution for tree-based models \cite{ribeiro2016lime,lundberg2017shap,lundberg2020trees}. Feature-attribution methods are nevertheless often used without placing their runtime on the same scale as prediction. In a constrained IoT setting, generating an explanation may require substantially more time and memory than producing the underlying class label. A detector that is inexpensive at inference time may become unsuitable if every prediction triggers a costly explanation. Predictive efficiency and explanation efficiency must consequently be measured separately.

Explanation availability also does not establish explanation reliability or operational value. Attribution sensitivity and adversarial manipulation have been demonstrated across post-hoc explanation settings \cite{ghorbani2019fragile,yeh2019infidelity,slack2020fooling}. An attribution may change under a small perturbation even when the model retains the same predicted class. Conversely, a stable attribution does not prove that the model relies on causal or semantically appropriate features. The usefulness of an explanation therefore depends on its computational cost, its behavior around the evaluated observation, and the decision context in which it is requested. These considerations motivate selective policies that reserve explanations for alerts, uncertain decisions, or other cases with a defined security priority.

This study develops that argument using CICIoT2023 \cite{neto2023ciciot} and a binary benign-versus-attack task. Logistic Regression implemented with the logistic-loss \texttt{SGDClassifier}, Decision Tree, Random Forest, and XGBoost models are evaluated on a leakage-safe corpus constructed after non-finite-value removal, exact-feature deduplication, and conservative label-collision handling. Predictive behavior is assessed on both the natural cleaned test distribution and a balanced sensitivity test. The natural distribution represents the prevalence retained after corpus construction, whereas the balanced test reveals how prevalence-sensitive metrics change when benign and attack observations are equally represented.

The analysis proceeds in stages. It first establishes whether the models provide adequate attack detection and false-alarm control, then examines training cost, model size, and the integrity of a feature that receives unusually high native importance. It subsequently measures TreeSHAP cost, explanation stability under prediction-preserving perturbations, and selective-explanation cost--coverage trade-offs. The purpose is not to argue that feature attribution makes an IDS trustworthy, but to determine whether explanatory analysis can be made computationally and operationally defensible in a resource-constrained setting.
\section{Related Work and Research Gap}

\subsection{IoT Intrusion Detection and CICIoT2023}

Machine-learning IDS research spans signature, anomaly, and hybrid detection, with model choice, feature representation, class imbalance, validation strategy, and deployment location all affecting the meaning of reported performance \cite{khraisat2019survey,khraisat2021iotids}. These dependencies are especially consequential in IoT networks, where heterogeneous devices and constrained monitoring points create different traffic and resource conditions from conventional enterprise networks \cite{khraisat2021iotids,moustafa2023explainable}. Consequently, a high score on a benchmark distribution is evidence about a particular experimental setting rather than direct evidence of feasibility on an operational gateway.

CICIoT2023 was constructed as a large-scale real-time IoT benchmark using a topology of 105 devices and an attack taxonomy covering 33 attacks in seven broad categories \cite{neto2023ciciot}. Its scale and family diversity have made it useful for binary and multiclass IDS experiments, but the authoritative benchmark description does not remove the need for study-specific audits of duplicates, label consistency, partition overlap, or prevalence effects. The present study therefore uses CICIoT2023 as its sole primary dataset while retaining original labels and attack families for diagnostics and constructing a binary target only after auditing the 39-feature representation.

Prior explainable IoT IDS research has proposed deep-learning frameworks, feature-attribution analyses, and surveys of opportunities for combining XAI with cyber defence \cite{houda2022trustids,keshk2023framework,moustafa2023explainable}. These studies establish the relevance of explanations to IoT security analysis, but they also show that predictive evaluation, explanatory analysis, and deployment claims must be separated. The contribution here is not another classifier benchmark alone; it is an evaluation of the additional computational and operational consequences that arise when explanations are requested from otherwise competitive detectors.

\subsection{Explainable AI for Intrusion Detection}

LIME explains an individual prediction through a locally fitted surrogate model, whereas SHAP expresses a prediction through Shapley-value-based feature contributions \cite{ribeiro2016lime,lundberg2017shap}. TreeSHAP provides algorithms specialized for decision trees and tree ensembles and supports local additive explanations that can be aggregated for broader model analysis \cite{lundberg2020trees}. These methods describe associations within a fitted predictor; their outputs do not by themselves establish causal relevance, semantic correctness, or analyst trust \cite{arrieta2020xai}.

Applying SHAP or LIME to an IDS is already well established. Le et al. applied SHAP to provide global and local explanations of Decision Tree and Random Forest intrusion-detection decisions on IoTID20, NF-BoT-IoT-v2, and NF-ToN-IoT-v2 \cite{le2022ensemble}. Patil et al. used LIME as the explanatory component of an IDS evaluated on CICIDS2017 with Decision Tree, Random Forest, SVM, and ensemble or voting models \cite{patil2022xaiids}. Abou El Houda et al. and Keshk et al. developed explainable deep-learning frameworks for IoT intrusion detection \cite{houda2022trustids,keshk2023framework}. More recent work has systematized model and explainer comparisons. XAI-IDS examines feature importance across multiple network datasets and models \cite{arreche2024xaiids}, while Wang et al. study SHAP- and LIME-oriented transparency for IoT intrusion detection \cite{wang2025iotxai}. This literature makes clear that the use of SHAP is not itself a novel contribution. The remaining operational question is how explanatory cost, local repeatability, and invocation policy interact under a defined deployment workload.

\subsection{Explanation Robustness, Stability, and Computational Cost}

Post-hoc explanations can change under small input changes even when predictive behavior changes little, motivating explicit sensitivity and robustness evaluation \cite{ghorbani2019fragile,yeh2019infidelity}. Explanation methods can also be manipulated or evaluated under adversarial conditions, further demonstrating that the presence of an explanation is not evidence of reliability \cite{slack2020fooling}. In the IDS domain, E-XAI evaluates black-box explanation methods using descriptive accuracy, sparsity, stability, efficiency, robustness, and completeness across three network-intrusion datasets \cite{arreche2024exai}. The gap addressed by the present study is therefore not the first consideration of XAI efficiency or stability in network intrusion detection.

Munilla and Khammas evaluate SHAP and LIME using DeepFool adversarial perturbations intended to stress the model decision boundary and alter model decisions \cite{munilla2026xairobustness}. The present study addresses a related but different construct. Its perturbations are bounded by validation-set statistics, restricted to selected continuous traffic features, accepted only when the predicted class remains unchanged, and required to satisfy an attack-confidence change no greater than 0.05. The two protocols consequently examine adversarial explanation robustness and local prediction-preserving explanation stability, respectively.

The TreeSHAP literature establishes efficient model-specific computation relative to generic Shapley-value estimation \cite{lundberg2020trees}, but algorithmic efficiency does not determine the wall-clock burden of a particular trained tree architecture, batch size, or deployment workload. Likewise, broad XAI evaluation frameworks can include efficiency without measuring explanation latency relative to the underlying prediction across operational batch sizes. This study therefore reports absolute explanation time, per-sample latency, throughput, and the explanation-to-prediction ratio together. It additionally connects these measurements to selective invocation. Selective classification provides a methodological precedent for risk--coverage trade-offs through abstention \cite{geifman2017selective}. The present work transfers this resource-allocation principle to explanation invocation while leaving the IDS classification output unchanged.

\subsection{Dataset Duplication, Leakage, and Evaluation Methodology}

Redundant records are a longstanding concern in IDS benchmarking. The analysis of KDD Cup 99 by Tavallaee et al. showed that extensive duplication could bias learning and evaluation, and the derived NSL-KDD benchmark was designed in part to reduce this redundancy \cite{tavallaee2009kdd}. More generally, IDS surveys identify dataset construction, validation strategy, and realistic evaluation as recurring methodological limitations \cite{khraisat2019survey,khraisat2021iotids}. The present work does not claim to discover the general duplicate-data problem. It provides a dataset-specific audit of CICIoT2023 and shows that 54.8176\% of valid rows were exact duplicate instances in the evaluated 39-feature representation.

The audit also distinguishes three label levels: the original label, the attack family, and the binary target. Attack-to-attack original-label collisions remain consistent after mapping to Attack~=~1, whereas benign--attack collisions do not. All cross-original-label hashes were nevertheless removed as a conservative semantic-consistency decision before deduplication. Deterministic hash-level partitioning then produced zero exact 39-feature-hash overlap among training, validation, and test partitions. This control is narrower and more precise than claiming that every form of dependence or leakage has been eliminated.

Table~\ref{tab:related-gap} summarizes the resulting research gap. The gap addressed by the present study is not the first consideration of XAI efficiency or stability in network intrusion detection. Rather, it is the joint operational evaluation of measured batch-dependent TreeSHAP cost relative to prediction, prediction-preserving local attribution stability, validation-calibrated selective explanation, prevalence-sensitive cost--coverage trade-offs, and exact-feature-hash-aware corpus construction.

\begin{table}[htbp]
  \centering
  \caption{Comparison of representative explainable IDS studies and the present work. ``Not explicitly evaluated'' indicates that the item was not a stated empirical focus of the cited study.}
  \label{tab:related-gap}
  \scriptsize
  \setlength{\tabcolsep}{0.5pt}
  \renewcommand{\arraystretch}{1.15}
  \begin{tabularx}{\textwidth}{>{\raggedright\arraybackslash}p{0.13\textwidth}>{\raggedright\arraybackslash}p{0.13\textwidth}>{\raggedright\arraybackslash}p{0.12\textwidth}>{\raggedright\arraybackslash}p{0.09\textwidth}*{4}{>{\centering\arraybackslash}X}}
    \toprule
    Study & Dataset(s) & Detection models & XAI method & Explicit XAI cost? & Stability? & Selective XAI? & Hash-aware split? \\
    \midrule
    Le et al. \cite{le2022ensemble} & IoTID20; NF-BoT-IoT-v2; NF-ToN-IoT-v2 & DT/RF ensemble-tree approach & SHAP & Not explicitly evaluated & Not explicitly evaluated & Not explicitly evaluated & Not explicitly evaluated \\
    Patil et al. \cite{patil2022xaiids} & CICIDS2017 & DT; RF; SVM; voting ensemble & LIME & Not explicitly evaluated & Not explicitly evaluated & Not explicitly evaluated & Not explicitly evaluated \\
    E-XAI \cite{arreche2024exai} & CICIDS2017; NSL-KDD; RoEduNet-\newline SIMARGL\linebreak 2021 & Multiple black-box models & SHAP; LIME & Evaluated & Evaluated & Not explicitly evaluated & Not explicitly evaluated \\
    Wang et al. \cite{wang2025iotxai} & IoT IDS datasets & Multiple ML models & SHAP; LIME & Not explicitly evaluated & Not explicitly evaluated & Not explicitly evaluated & Not explicitly evaluated \\
    Munilla and Khammas \cite{munilla2026xairobustness} & BoT-IoT; Edge-IIoT; N-BaIoT & CNN; DNN; LSTM; RF & SHAP; LIME & Not explicitly evaluated & Evaluated & Not explicitly evaluated & Not explicitly evaluated \\
    Present study & CICIoT2023 & LR; DT; RF; XGBoost & \shortstack{Tree\\SHAP} & Yes & Evaluated & Yes & Yes \\
    \bottomrule
  \end{tabularx}
\end{table}
\section{Research Motivation and Contributions}

The immediate methodological motivation arises from the structure of CICIoT2023. The raw corpus contains extensive exact duplication. If rows were assigned independently to training and test partitions, identical 39-dimensional feature vectors could appear on both sides of the evaluation boundary. Performance measured under such a split could reflect exact feature-vector reuse rather than generalization to unseen observations. Label collisions create an additional annotation-consistency concern because some identical feature vectors are associated with different original labels or attack families. A rigorous predictive comparison must address these issues before model performance is interpreted.

A second motivation is the dependence of commonly reported metrics on the test distribution. The cleaned natural test set remains strongly attack dominated. Performance on this distribution is relevant to the constructed corpus, but it does not by itself show how a model behaves when benign traffic receives equal representation. Reporting a balanced sensitivity test alongside the natural test makes this dependence explicit without replacing the primary test distribution or merging CICIoT2023 with an external dataset.

The third motivation concerns the gap between prediction and explanation. The resource requirements of a classifier do not determine the cost of its post-hoc explanations. Moreover, neither native feature importance nor a single attribution plot establishes that a feature is indispensable to prediction or that an explanation is reliable. The targeted audit and ablation of \texttt{Number} illustrate why feature integrity and predictive dependence must be examined before explanation outputs are given operational meaning.

This work makes four connected contributions. First, it constructs a leakage-safe binary CICIoT2023 corpus through non-finite-value handling, exact-feature deduplication, conservative original-label collision removal, and deterministic hash-level partitioning. Second, it compares Logistic Regression, Decision Tree, Random Forest, and XGBoost under natural and balanced test distributions, emphasizing recall, F1, false-positive rate, and PR-AUC rather than treating accuracy as the primary criterion. Third, it measures prediction and model efficiency and audits the unusually influential \texttt{Number} feature through targeted ablation. Fourth, it empirically measures TreeSHAP explanation cost, explanation stability under prediction-preserving perturbations, and selective-explanation cost--coverage trade-offs.

The resulting manuscript is organized around trade-offs rather than a universal model ranking. A model may provide the highest F1 score without producing the lowest false-positive rate; a compact single tree may remain competitive with larger ensembles; and an efficient predictor may still incur prohibitive explanation overhead. Separating these dimensions is necessary to move from detection accuracy toward operationally useful explainable intrusion detection.

The central question is therefore: when predictive performance is similar, what is the computational and operational cost of obtaining explanations, and can explanation workload be reduced without losing security-relevant explanatory coverage?

\section{Research Questions}

The study addresses four research questions that follow the progression from predictive performance to operational explanation.

\textbf{RQ1: Predictive effectiveness.} How do Logistic Regression, Decision Tree, Random Forest, and XGBoost compare on the binary CICIoT2023 task when recall, precision, F1 score, false-positive rate, and PR-AUC are prioritized over accuracy?

\textbf{RQ2: Explanation cost.} How much latency and throughput overhead does TreeSHAP introduce relative to prediction alone for the tree-based detectors, and how does this cost vary across model architecture and batch size?

\textbf{RQ3: Explanation stability.} How stable are local feature attributions under small prediction-preserving perturbations, and how does stability vary across model architecture, perturbation strength, and benign-versus-attack observations?

\textbf{RQ4: Selective explanation.} To what extent can security-aware or uncertainty-aware selection policies reduce explanation workload while retaining the decisions most relevant to operational review?

RQ1--RQ4 are answered by the empirical study. Keeping the questions separate prevents predictive effectiveness from being used as a proxy for explanation efficiency, stability, or utility.

\section{Experimental Methodology}

\subsection{CICIoT2023 Dataset}

CICIoT2023 is the sole dataset used in the present study \cite{neto2023ciciot}. The benchmark was constructed around 105 IoT devices and an attack taxonomy containing 33 attack types \cite{neto2023ciciot}. The raw corpus analyzed here contains 46,776,700 observations and 39 features. It includes benign traffic and attacks belonging to the DDoS, DoS, Recon, Web, BruteForce, Spoofing, and Mirai families. The primary classification task assigns benign observations to class 0 and all attacks to class 1. Original attack labels and family assignments are retained for descriptive and diagnostic analysis even though the fitted models address the binary task.

Table~\ref{tab:raw-family-distribution} reports the corrected family distribution in the raw corpus. DDoS and DoS traffic account for most observations, while benign traffic constitutes a much smaller proportion. This distribution motivates the use of class-sensitive metrics and a separate balanced sensitivity evaluation. No external dataset was merged with CICIoT2023, and cross-dataset validation is outside the present experimental scope.

\begin{table}[htbp]
  \centering
  \caption{Corrected family distribution in the raw CICIoT2023 corpus.}
  \label{tab:raw-family-distribution}
  \small
  \begin{tabular}{lr@{\hspace{1.2cm}}lr}
    \toprule
    Family & Observations & Family & Observations \\
    \midrule
    DDoS       & 33,984,450 & DoS        & 7,845,120 \\
    Mirai      & 2,634,054  & Benign     & 1,098,191 \\
    Recon      & 690,534    & Spoofing   & 486,458 \\
    Web        & 24,829     & BruteForce & 13,064 \\
    \bottomrule
  \end{tabular}
\end{table}

\subsection{Data-Quality Audit}

The data-quality audit covered all 309 CSV files. It examined schema consistency and counted missing and infinite values before any model training or partitioning. Across the 46,776,700 raw rows, the audit identified 1,449 missing-value cells and 1,037 infinite-value cells. These are cell counts rather than row counts, and more than one invalid cell could occur in a single record. The non-finite cells were concentrated in 1,040 rows, all of which were removed. This operation left 46,775,660 valid observations. No schema inconsistency was found across the audited files.

The distinction between invalid cells and removed rows is retained in the reporting because adding the missing- and infinite-cell counts would overstate the number of excluded observations. Removal was performed before hashing so that duplicate and collision statistics were computed only over finite 39-dimensional feature vectors.

Table~\ref{tab:data-quality} summarizes the file-level and observation-level data-quality audit.

\begin{table}[htbp]
  \centering
  \caption{Data-quality audit of the raw CICIoT2023 files.}
  \label{tab:data-quality}
  \small
  \begin{tabular}{lr}
    \toprule
    Audit quantity & Count \\
    \midrule
    CSV files audited & 309 \\
    Raw rows & 46,776,700 \\
    Missing-value cells & 1,449 \\
    Infinite-value cells & 1,037 \\
    Rows removed for non-finite values & 1,040 \\
    Schema inconsistencies & 0 \\
    Valid rows after removal & 46,775,660 \\
    \bottomrule
  \end{tabular}
\end{table}

\subsection{Duplicate and Label-Collision Analysis}

An exact hash was computed over the complete 39-feature vector of every valid observation. The 46,775,660 valid rows contained 21,134,351 unique feature vectors, 1,893,344 duplicate groups, and 25,641,309 duplicate instances. The resulting exact duplicate rate was 54.8176\%. This rate is sufficiently high that a na\"ive random row-level split could place identical feature vectors in training and test data, allowing exact feature-vector leakage and potentially optimistic performance estimates.

Duplicate feature vectors were also audited for label consistency. The analysis identified 533,528 hashes associated with more than one original label, 451,969 hashes crossing attack-family boundaries, and 905 hashes associated with both benign and attack observations. These categories are related and should not be summed as if they were disjoint. Benign-versus-attack collisions are inconsistent at the binary-target level because the same feature vector maps to both class 0 and class 1. In contrast, attack-to-attack original-label collisions remain binary-target consistent because every involved attack label maps to Attack = 1. The preprocessing procedure nevertheless excluded all cross-original-label hashes as a conservative methodological choice. This decision preserved semantic label consistency in the clean primary corpus and avoided retaining feature-identical observations carrying conflicting source annotations. After collision handling, 20,600,823 safe unique hashes remained.

Table~\ref{tab:duplicate-audit} reports the duplicate and label-collision counts.

\begin{table}[htbp]
  \centering
  \caption{Exact-duplicate and label-collision audit.}
  \label{tab:duplicate-audit}
  \small
  \begin{tabular}{lr}
    \toprule
    Audit quantity & Count \\
    \midrule
    Valid rows & 46,775,660 \\
    Unique 39-dimensional feature vectors & 21,134,351 \\
    Duplicate groups & 1,893,344 \\
    Duplicate instances & 25,641,309 \\
    Exact duplicate rate & 54.8176\% \\
    Cross-original-label collision hashes & 533,528 \\
    Cross-family collision hashes & 451,969 \\
    Benign-versus-attack collision hashes & 905 \\
    Safe unique hashes after collision handling & 20,600,823 \\
    \bottomrule
  \end{tabular}
\end{table}

\subsection{Leakage-Safe Corpus Construction}

The clean corpus was constructed in five ordered steps. Rows containing a missing or infinite value were removed. An exact hash was then computed over each 39-feature vector. As the conservative semantic-consistency choice described above, hashes associated with conflicting original labels were excluded, and the remaining exact duplicates were collapsed to one observation per feature vector. Finally, each surviving feature hash was assigned deterministically to a training, validation, or test partition.

The resulting corpus contained 20,600,823 unique observations. The nominal split was 70\% training, 15\% validation, and 15\% test. Deterministic hash assignment produced 14,421,214 training observations, 3,089,510 validation observations, and 3,090,099 natural-test observations. The feature-hash overlap was zero for train versus validation, train versus test, and validation versus test. Exact feature-vector leakage across the three partitions was therefore eliminated.

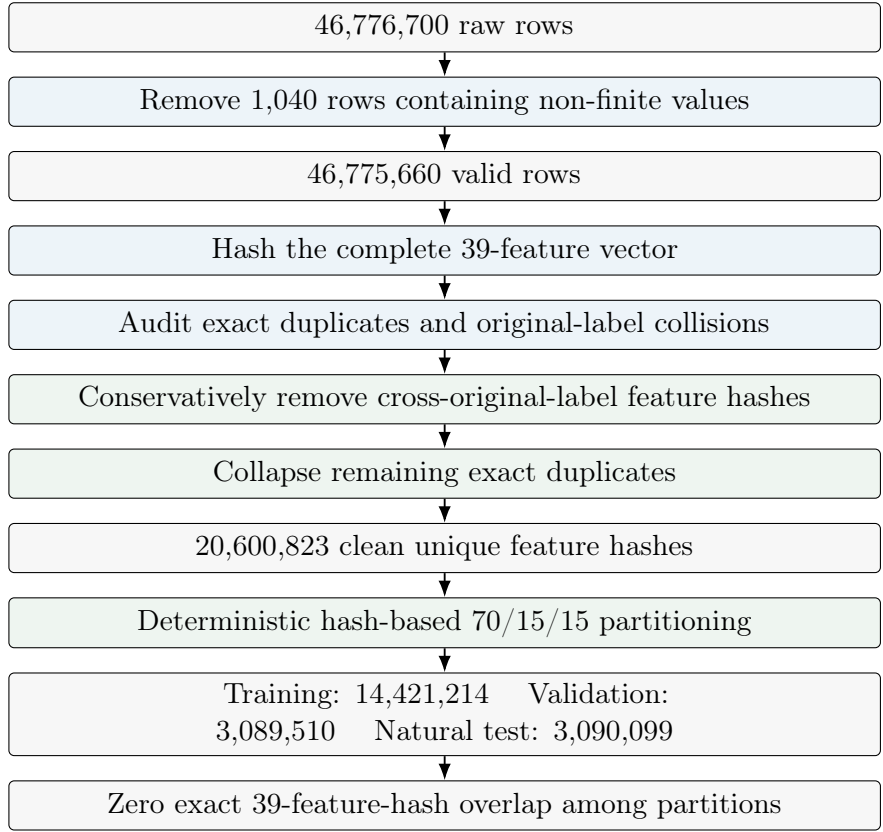
\begin{figure}[htbp]
  \centering
  \begin{tikzpicture}[
    node distance=3.2mm,
    stage/.style={draw,rounded corners=2pt,align=center,text width=0.82\textwidth,minimum height=6.5mm,fill=gray!6,font=\small},
    audit/.style={stage,fill=dtcolor!8},
    clean/.style={stage,fill=xgbcolor!8},
    arrow/.style={-{Latex[length=2.2mm]},line width=0.7pt}
  ]
    \node[stage] (raw) {46,776,700 raw rows};
    \node[audit,below=of raw] (finite) {Remove 1,040 rows containing non-finite values};
    \node[stage,below=of finite] (valid) {46,775,660 valid rows};
    \node[audit,below=of valid] (hash) {Hash the complete 39-feature vector};
    \node[audit,below=of hash] (audit) {Audit exact duplicates and original-label collisions};
    \node[clean,below=of audit] (remove) {Conservatively remove cross-original-label feature hashes};
    \node[clean,below=of remove] (collapse) {Collapse remaining exact duplicates};
    \node[stage,below=of collapse] (corpus) {20,600,823 clean unique feature hashes};
    \node[clean,below=of corpus] (split) {Deterministic hash-based 70/15/15 partitioning};
    \node[stage,below=of split] (parts) {Training: 14,421,214 \quad Validation: 3,089,510 \quad Natural test: 3,090,099};
    \node[stage,below=of parts] (overlap) {Zero exact 39-feature-hash overlap among partitions};

    \foreach \a/\b in {raw/finite,finite/valid,valid/hash,hash/audit,audit/remove,remove/collapse,collapse/corpus,corpus/split,split/parts,parts/overlap}
      \draw[arrow] (\a) -- (\b);
  \end{tikzpicture}
  \caption{Leakage-aware CICIoT2023 corpus-construction workflow. The procedure removes non-finite observations, conservatively excludes cross-original-label feature hashes, collapses exact duplicates, and partitions the surviving unique feature vectors deterministically.}
  \label{fig:data-construction}
\end{figure}

Figure~\ref{fig:data-construction} summarizes the ordered construction and partitioning procedure.

Table~\ref{tab:data-splits} reports the resulting partition sizes and class composition.

\begin{table}[htbp]
  \centering
  \caption{Class composition of the leakage-safe corpus and balanced sensitivity test.}
  \label{tab:data-splits}
  \small
  \begin{tabular}{lrrr}
    \toprule
    Partition & Total & Benign & Attack \\
    \midrule
    Training (70\%) & 14,421,214 & 765,187 & 13,656,027 \\
    Validation (15\%) & 3,089,510 & 164,271 & 2,925,239 \\
    Natural test (15\%) & 3,090,099 & 163,513 & 2,926,586 \\
    Balanced sensitivity test & 327,026 & 163,513 & 163,513 \\
    \bottomrule
  \end{tabular}
\end{table}

This procedure eliminates exact overlap, but it does not imply that all statistical dependence among related non-identical flows has been removed. The evaluation is therefore described specifically as leakage-safe with respect to exact 39-feature hashes.

\subsection{Predictive Models}

The model set represents four levels of functional and structural complexity. Logistic Regression provides a lightweight linear baseline and was implemented with \texttt{SGDClassifier(loss="log\_loss")} in scikit-learn \cite{pedregosa2011scikit}. It was trained for five epochs on all 14,421,214 training observations. A \texttt{StandardScaler} was fitted on the training data only and then applied without refitting to the validation and test sets. Balanced class weights were 9.4233266 for benign traffic and 0.5280165 for attack traffic. The principal classification threshold was 0.50. A threshold of 0.11 selected through validation-set F1 optimization was retained only for sensitivity analysis.

Decision Tree, Random Forest, and XGBoost were trained on a common deterministic sample of 5,000,000 observations from the training partition. Holding this sample constant supports direct comparison among the three nonlinear models. The Decision Tree used \texttt{max\_depth=20}, \texttt{min\_samples\_leaf=20}, and balanced class weights. The Random Forest \cite{breiman2001random} contained 150 estimators, with \texttt{max\_depth=20}, \texttt{min\_samples\_leaf=20}, the square root of the feature count considered at each split, and balanced-subsample class weights.

XGBoost \cite{chen2016xgboost} used 300 estimators, a maximum depth of 8, a learning rate of 0.10, a row-subsampling proportion of 0.80, a column-subsampling proportion of 0.80, a minimum child weight of 5, and the histogram tree method. Balanced sample weighting was derived from the full training distribution. The global experiment seed was 2026. The test partitions were not used for threshold selection or model fitting.

\subsection{Experimental Environment}

Experiments were executed on a Microsoft Windows~11 Pro host equipped with an Intel(R) Core(TM)~7 240H processor, 10 physical cores, 16 logical processors, and 31.71~GB of system RAM. The operating-system version was 10.0.26200, build 26200. The Python version was 3.12.0. Principal package versions were NumPy~2.4.6, pandas~3.0.5, scikit-learn~1.9.0, XGBoost~3.4.0, SHAP~0.52.0, joblib~1.5.3, psutil~7.2.2, and PyArrow~25.0.0. The global random seed was 2026.

No explicit values were assigned to \texttt{OMP\_NUM\_THREADS}, \texttt{MKL\_NUM\_THREADS}, \texttt{OPENBLAS\_NUM\_THREADS}, or \texttt{NUMEXPR\_NUM\_THREADS}. This observation does not imply that the underlying libraries used a specific fixed thread count. All reported training, inference, and explanation timings are wall-clock measurements obtained in this hardware and software environment. They are environment dependent and must not be interpreted as direct measurements on a constrained edge device.

\subsection{Reproducibility and Experimental Controls}

Reproducibility controls were applied across data construction, model evaluation, explanation timing, stability analysis, and selective-policy calibration. The corpus split was deterministic at the complete 39-feature-hash level, and the global experiment seed was fixed at 2026. Decision Tree, Random Forest, and XGBoost used the same deterministic 5,000,000-observation training sample. The natural and balanced evaluations reused the same fitted models and fixed decision thresholds; neither distribution initiated retraining.

Selective-explanation thresholds were calibrated only from validation false negatives, and test labels were never used to choose a threshold. TreeSHAP cost used one fixed balanced master sample of 5,000 observations, with the prediction and adaptive explanation repeat schedules stated in Section~\ref{sec:explanation-cost}. Perturbation geometry in the stability experiment was derived from validation-set interquartile ranges and percentile bounds. Full-test selective-explanation costs are identified as throughput-based projections rather than direct runtime measurements. The deterministic seeds, split construction, model configurations, evaluation protocols, and software environment are documented to support reproducibility. The experimental scripts, trained model artifacts, derived result files, and reproducibility documentation supporting this study are publicly available through the project repository and its archived Zenodo release (DOI: 10.5281/zenodo.21879038).

\subsection{Evaluation Metrics}

Let true positives (TP) denote attacks classified as attacks, false negatives (FN) denote missed attacks, false positives (FP) denote benign observations classified as attacks, and true negatives (TN) denote correctly classified benign observations. The principal threshold-dependent metrics are
\begin{align}
  \mathrm{Precision} &= \frac{\mathrm{TP}}{\mathrm{TP}+\mathrm{FP}}, \\
  \mathrm{Recall} &= \frac{\mathrm{TP}}{\mathrm{TP}+\mathrm{FN}}, \\
  \mathrm{F1} &= 2\frac{\mathrm{Precision}\,\mathrm{Recall}}
  {\mathrm{Precision}+\mathrm{Recall}}, \\
  \mathrm{FPR} &= \frac{\mathrm{FP}}{\mathrm{FP}+\mathrm{TN}}.
\end{align}

PR-AUC is used as the principal threshold-independent summary because it characterizes the precision--recall trade-off and remains directly connected to attack-detection performance. Accuracy and receiver-operating-characteristic area under the curve (ROC-AUC) are reported as secondary metrics. Accuracy is not used as the primary basis for model selection because the cleaned corpus remains severely imbalanced. False-positive rate receives separate attention because benign false alarms create an operational burden that may not be visible in an aggregate F1 or accuracy value.

Computational measurements comprise training time, inference latency, and serialized model size. Prediction cost and explanation cost are treated separately. The TreeSHAP experiment measured prediction latency, explanation latency, prediction and explanation throughput, observed process-memory effects, and the explanation-to-inference overhead ratio at batch sizes of 1, 10, 100, 1,000, and 5,000.

\subsection{Natural and Balanced Evaluation Protocol}

The natural test set contains 3,090,099 observations: 163,513 benign and 2,926,586 attack observations. It preserves the attack-dominated prevalence of the cleaned corpus and is the principal test distribution. A balanced sensitivity test was constructed using all 163,513 benign test observations and 163,513 attack observations, for a total of 327,026. The balanced test does not supplant the natural test and is not used for tuning. It provides a controlled view of metric sensitivity when the two binary classes have equal representation.

Both tests use the same fitted models and fixed classification thresholds. Consequently, changes in precision, F1, or accuracy across the two tests reflect the evaluation distribution rather than retraining. The false-positive rate remains directly comparable because the benign observations are unchanged. PR-AUC is reported separately for each distribution because precision depends on prevalence. Results are interpreted jointly: the natural test represents performance on the cleaned corpus distribution, while the balanced test exposes conclusions that depend strongly on attack prevalence.
\section{Predictive Results}

Tables~\ref{tab:natural-results} and \ref{tab:balanced-results} report predictive performance on the natural and balanced test distributions, respectively.

\begin{table}[htbp]
  \centering
  \caption{Binary detection performance on the natural test set. Values are expressed as proportions.}
  \label{tab:natural-results}
  \scriptsize
  \setlength{\tabcolsep}{3.5pt}
  \begin{tabular}{lccccccc}
    \toprule
    Model & Accuracy & Precision & Recall & F1 & FPR & ROC-AUC & PR-AUC \\
    \midrule
    Logistic Regression & 0.821520 & 0.979635 & 0.828777 & 0.897914 & 0.308367 & 0.816803 & 0.987239 \\
    Decision Tree & 0.972772 & 0.999442 & 0.971794 & 0.985424 & 0.009718 & 0.992142 & 0.999379 \\
    Random Forest & 0.970275 & 0.999869 & 0.968741 & 0.984059 & 0.002269 & 0.995555 & 0.999755 \\
    XGBoost & 0.973098 & 0.999735 & 0.971852 & 0.985597 & 0.004605 & 0.996212 & 0.999791 \\
    \bottomrule
  \end{tabular}
\end{table}

\begin{table}[htbp]
  \centering
  \caption{Binary detection performance on the balanced sensitivity test. Values are expressed as proportions.}
  \label{tab:balanced-results}
  \scriptsize
  \setlength{\tabcolsep}{3.5pt}
  \begin{tabular}{lccccccc}
    \toprule
    Model & Accuracy & Precision & Recall & F1 & FPR & ROC-AUC & PR-AUC \\
    \midrule
    Logistic Regression & 0.760692 & 0.729055 & 0.829751 & 0.776151 & 0.308367 & 0.817303 & 0.840948 \\
    Decision Tree & 0.981228 & 0.990103 & 0.972173 & 0.981056 & 0.009718 & 0.992277 & 0.990255 \\
    Random Forest & 0.983533 & 0.997665 & 0.969336 & 0.983296 & 0.002269 & 0.995602 & 0.996670 \\
    XGBoost & 0.983842 & 0.995286 & 0.972290 & 0.983653 & 0.004605 & 0.996248 & 0.997126 \\
    \bottomrule
  \end{tabular}
\end{table}

\subsection{Linear vs Nonlinear Models}

The difference between Logistic Regression and the three tree-based models is visible on both test distributions. At the 0.50 threshold, Logistic Regression attained 0.828777 recall and 0.897914 F1 on the natural test, but its false-positive rate was 0.308367. Its accuracy of 0.821520 therefore does not fully describe the operational weakness of the classifier: almost one third of benign observations were classified as attacks. The nonlinear models reduced the false-positive rate to 0.009718 or below while producing F1 scores above 0.984 on the same test distribution.

The balanced evaluation makes the limitation of the linear model more explicit. Logistic Regression retained 0.829751 recall, but precision declined to 0.729055 and F1 to 0.776151. The false-positive rate remained 0.308367 because the benign evaluation set was unchanged. By contrast, the nonlinear models produced balanced-test F1 scores from 0.981056 to 0.983653. The observation is statistical rather than causal: the results show that the fitted linear decision function separated the evaluated traffic less effectively, but they do not identify why individual features interact nonlinearly.

Threshold selection further illustrates the danger of relying on a single aggregate objective. Validation-based F1 optimization selected a Logistic Regression threshold of 0.11. On the natural test, this threshold produced 0.887032 accuracy, 0.959895 precision, 0.919122 recall, 0.939066 F1, and 0.687321 FPR. On the balanced test, it produced 0.616067 accuracy, 0.572236 precision, 0.919456 recall, 0.705435 F1, and 0.687321 FPR. ROC-AUC and PR-AUC are threshold independent: they remained 0.816803 and 0.987239 on the natural test and 0.817303 and 0.840948 on the balanced test. Although the lower threshold improved natural-test recall and F1, it produced an operationally unacceptable increase in benign false positives. The 0.50 threshold is therefore retained as the principal baseline, and the 0.11 threshold is reported only as sensitivity analysis.

\subsection{Ensemble Model Comparison}

The three tree-based models were close in aggregate predictive performance, but their error profiles were not identical. XGBoost achieved the highest natural-test F1, 0.985597, and the highest balanced-test F1, 0.983653. It also produced the highest ROC-AUC and PR-AUC on both distributions: 0.996212 and 0.999791 on the natural test, and 0.996248 and 0.997126 on the balanced test. These observed metrics give XGBoost the strongest overall predictive profile among the evaluated models. However, the differences among the tree-based models are small and are interpreted descriptively rather than as evidence of statistical superiority.

Random Forest produced the lowest false-positive rate, 0.002269, on both tests. Its natural-test precision was 0.999869 and its balanced-test precision was 0.997665, both higher than the corresponding XGBoost values. This false-alarm advantage was accompanied by lower recall: 0.968741 on the natural test and 0.969336 on the balanced test. A deployment that places greater cost on benign false alarms could therefore prefer Random Forest even though its F1 is slightly lower.

Decision Tree remained notably competitive despite its simpler structure. Its natural-test F1 of 0.985424 was close to XGBoost's 0.985597, and its recall was 0.971794 compared with 0.971852 for XGBoost. On the balanced test, Decision Tree produced 0.981056 F1 and a 0.009718 false-positive rate. The single tree does not match the ensembles on every measure, but its performance is sufficiently close that model complexity and resource cost become relevant selection criteria. Detection accuracy alone does not resolve this trade-off.

The distribution comparison also changes the interpretation of apparently high values. Precision and accuracy decrease or increase as class prevalence changes, even though each model and threshold remain fixed. PR-AUC likewise changes between the natural and balanced tests. These shifts reinforce the need to report the test distribution alongside every metric and to avoid generalizing a result beyond the evaluated traffic composition.

\subsection{Efficiency Trade-offs}

Table~\ref{tab:efficiency-results} reports the completed training-time and serialized-size measurements. The comparison between Decision Tree, Random Forest, and XGBoost is based on the same 5,000,000-observation sample. Logistic Regression used all 14,421,214 training observations and five epochs, so its 38.14-second training time is not directly comparable with the common-sample tree results without accounting for the different training workloads.

\begin{table}[htbp]
  \centering
  \caption{Recorded training time and serialized classifier-model size.}
  \label{tab:efficiency-results}
  \small
  \begin{adjustbox}{max width=\textwidth}
  \begin{tabular}{lrrr}
    \toprule
    Model & Training observations & Training time (s) & Model size (MB) \\
    \midrule
    Logistic Regression & 14,421,214 & 38.14 & 0.002238 \\
    Decision Tree & 5,000,000 & 13.72 & $\approx0.498$ \\
    Random Forest & 5,000,000 & 99.28 & 22.466 \\
    XGBoost & 5,000,000 & 35.02 & 0.895 \\
    \bottomrule
  \end{tabular}
  \end{adjustbox}
  \par\vspace{2pt}
  \begin{minipage}{0.96\textwidth}
    \footnotesize\textit{Note:} The separately serialized StandardScaler occupied 0.001433~MB; the classifier and scaler together occupied approximately 0.003671~MB.
  \end{minipage}
\end{table}

Decision Tree required 13.72 seconds to train and approximately 0.498 MB of serialized storage on the experimental host. Its combination of compact serialized size, lower host training time, and F1 close to the ensemble models makes it a candidate for further evaluation in resource-constrained deployment settings rather than merely a weak baseline. These measurements do not demonstrate edge-device feasibility. Random Forest required 99.28 seconds and 22.466 MB, the largest host training time and model size among the three models trained on the common sample. The additional storage and training cost accompany the lowest observed false-positive rate, showing that efficiency and security error costs can point toward different choices.

XGBoost trained in 35.02 seconds on the experimental host and occupied 0.895 MB. Given its leading F1, ROC-AUC, and PR-AUC values, these measurements support the interpretation that XGBoost provides the strongest predictive/host-efficiency compromise among the completed experiments. This is not a universal ranking or evidence of edge-device feasibility. Decision Tree is smaller and faster to train, while Random Forest has the lowest false-positive rate. Moreover, serialized size and host training time do not establish inference cost, and none of these measurements establishes explanation cost. The Stage-5 latency and TreeSHAP benchmark is therefore necessary before deployment efficiency can be assessed, and actual constrained-hardware evaluation would still be required.

\subsection{Feature-Integrity Analysis}

Native XGBoost feature importance was heavily concentrated on \texttt{Number}. High native importance can arise because a feature provides useful split points, but it does not establish causal relevance, data leakage, or dependence that cannot be substituted by other features. A targeted integrity audit was therefore performed before using this importance pattern to motivate explanation claims.

The benign-versus-attack KS statistic for \texttt{Number} was 0.939147, the largest among the audited features. Other high values included 0.861050 for \texttt{HTTPS}, 0.806869 for \texttt{ack\_flag\_number}, 0.793179 for \texttt{IAT}, 0.780622 for \texttt{Header\_Length}, 0.775387 for \texttt{Rate}, and 0.735218 for \texttt{ack\_count}. These statistics show strong univariate distributional separation, but they do not identify whether the fitted model depends exclusively on any one feature.

Table~\ref{tab:ks-results} reports the KS statistics used in the targeted integrity audit.

\begin{table}[htbp]
  \centering
  \caption{Benign-versus-attack KS statistics for features examined in the targeted integrity audit.}
  \label{tab:ks-results}
  \small
  \begin{tabular}{lr@{\hspace{1.2cm}}lr}
    \toprule
    Feature & KS statistic & Feature & KS statistic \\
    \midrule
    \texttt{Number} & 0.939147 & \texttt{HTTPS} & 0.861050 \\
    \texttt{ack\_flag\_number} & 0.806869 & \texttt{IAT} & 0.793179 \\
    \texttt{Header\_Length} & 0.780622 & \texttt{Rate} & 0.775387 \\
    \texttt{ack\_count} & 0.735218 & & \\
    \bottomrule
  \end{tabular}
\end{table}
\FloatBarrier

Within the natural test set, \texttt{Number} had 99 unique values and an approximate unique-value ratio of 0.0032\%. Its most common value accounted for approximately 88.74\% of observations, and it was not flagged as high-uniqueness or identifier-like. The benign distribution had a mean of approximately 9.998 and a median of 10, whereas the attack distribution had a mean of approximately 94.40 and a median of 100. The feature therefore separates broad traffic regimes while taking a small set of repeated values. This evidence neither proves nor disproves leakage; it establishes the need for a direct dependence test.

\subsection{\texttt{Number} Feature Ablation}

Decision Tree and XGBoost were retrained after removing \texttt{Number}, leaving 38 features. Each ablated model used the same deterministic 5,000,000-observation training sample as its 39-feature counterpart, so the comparison isolates the removal of the feature within the recorded protocol.

Table~\ref{tab:number-ablation} compares the 39-feature and ablated configurations.

\begin{table}[htbp]
  \centering
  \caption{F1 and false-positive-rate results from the targeted \texttt{Number} ablation.}
  \label{tab:number-ablation}
  \small
  \begin{adjustbox}{max width=\textwidth}
  \begin{tabular}{lllrrr}
    \toprule
    Model & Test distribution & Metric & 39 features & Without \texttt{Number} & $\Delta$ \\
    \midrule
    XGBoost & Natural & F1 & 0.985597 & 0.985574 & $-0.000023$ \\
    XGBoost & Balanced & F1 & 0.983653 & 0.983625 & $-0.000028$ \\
    Decision Tree & Natural & F1 & 0.985424 & 0.984974 & $-0.000450$ \\
    Decision Tree & Balanced & F1 & 0.981056 & 0.981313 & $+0.000257$ \\
    Decision Tree & Both tests & FPR & 0.009718 & 0.008183 & $-0.001535$ \\
    \bottomrule
  \end{tabular}
  \end{adjustbox}
\end{table}

Removing \texttt{Number} changed XGBoost F1 by $-0.000023$ on the natural test and $-0.000028$ on the balanced test. Predictive performance was therefore virtually unchanged despite the feature's high native importance and strong class separation. Decision Tree natural-test F1 decreased from 0.985424 to 0.984974, a change of $-0.000450$, whereas balanced-test F1 increased from 0.981056 to 0.981313, a change of $+0.000257$. Its false-positive rate decreased from 0.009718 to 0.008183, corresponding to $\Delta\mathrm{FPR}=-0.001535$.

The ablation demonstrates that XGBoost's predictive effectiveness does not materially depend on \texttt{Number} alone. It also shows why native feature importance should not be equated with explanation quality or indispensable model dependence. More broadly, the result reinforces the central argument of this section: even after a detector achieves high predictive performance, model selection still requires evidence about resource cost and the behavior of the explanatory layer. The following sections therefore evaluate TreeSHAP cost, stability, and selective invocation relative to prediction alone.
\section{Explanation Cost}
\label{sec:explanation-cost}

TreeSHAP provides model-specific additive feature attributions for tree-based predictors \cite{lundberg2017shap,lundberg2020trees}. Its cost was evaluated here for Decision Tree, Random Forest, and XGBoost using a deterministic balanced master sample of 5,000 observations containing 2,500 benign and 2,500 attack cases, each represented by the same 39 features used for prediction. Measurements were collected at batch sizes of 1, 10, 100, 1,000, and 5,000. Prediction timing used 10 repetitions at every batch size. TreeSHAP timing used an adaptive schedule of 10 repetitions for batches 1 and 10, five for batch 100, three for batch 1,000, and one for batch 5,000.

For model $m$ and batch $b$, let $T^{\mathrm{predict}}_{m,b}$ denote prediction time and $T^{\mathrm{explain}}_{m,b}$ denote TreeSHAP time. The explanation-overhead ratio is
\begin{equation}
  O_{\mathrm{XAI}}(m,b)=
  \frac{T^{\mathrm{explain}}_{m,b}}
       {T^{\mathrm{predict}}_{m,b}},
\end{equation}
and the absolute incremental explanation time is
\begin{equation}
  C_{\mathrm{XAI}}(m,b)=
  T^{\mathrm{explain}}_{m,b}-T^{\mathrm{predict}}_{m,b}.
\end{equation}
An overhead ratio is informative only in conjunction with absolute latency. When prediction takes tens of microseconds, a moderate absolute explanation time can produce a very large ratio. The analysis therefore considers absolute explanation latency, latency per sample, explanation throughput, and the explanation-to-prediction ratio jointly.

\subsection{Scaling with Batch Size}

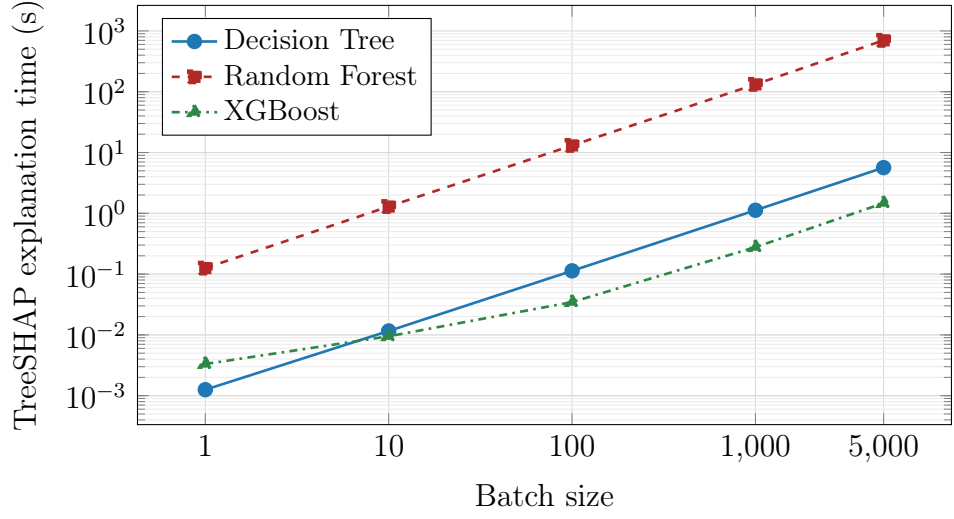
\begin{figure}[htbp]
  \centering
  \begin{tikzpicture}
    \begin{axis}[
      width=0.90\textwidth,
      height=0.52\textwidth,
      xmode=log,
      ymode=log,
      log basis x=10,
      log basis y=10,
      xlabel={Batch size},
      ylabel={TreeSHAP explanation time (s)},
      xtick={1,10,100,1000,5000},
      xticklabels={1,10,100,{1,000},{5,000}},
      grid=both,
      minor grid style={gray!15},
      major grid style={gray!30},
      legend pos=north west,
      legend cell align=left,
      legend style={font=\small}
    ]
      \addplot[color=dtcolor,mark=*,mark size=2.4pt,line width=1pt,solid] coordinates {
        (1,0.001255) (10,0.011603) (100,0.113108)
        (1000,1.123412) (5000,5.642689)};
      \addlegendentry{Decision Tree}
      \addplot[color=rfcolor,mark=square*,mark size=2.3pt,line width=1pt,dashed] coordinates {
        (1,0.125572) (10,1.293386) (100,13.077294)
        (1000,132.182032) (5000,700.758801)};
      \addlegendentry{Random Forest}
      \addplot[color=xgbcolor,mark=triangle*,mark size=2.7pt,line width=1pt,dashdotted] coordinates {
        (1,0.003321) (10,0.009487) (100,0.034625)
        (1000,0.277383) (5000,1.471088)};
      \addlegendentry{XGBoost}
    \end{axis}
  \end{tikzpicture}
  \caption{TreeSHAP explanation-time scaling on the experimental host. Both axes are logarithmic; marker shapes and line styles distinguish models independently of color.}
  \label{fig:cost-scaling}
\end{figure}

Figure~\ref{fig:cost-scaling} shows that TreeSHAP cost increased with batch size for all three architectures, but at markedly different rates and absolute levels. Decision Tree prediction medians were 0.000049, 0.000054, 0.000060, 0.000122, and 0.000433~s across batches 1, 10, 100, 1,000, and 5,000. The reported TreeSHAP times were 0.001255, 0.011603, 0.113108, 1.123412, and 5.642689~s, producing overhead ratios of 25.46$\times$, 216.06$\times$, 1,894.60$\times$, 9,196.98$\times$, and 13,025.60$\times$. Under the adaptive schedule, the 5,000-sample TreeSHAP value was one recorded measurement; the smaller-batch values summarized repeated timings. The large terminal ratio does not mean that Decision Tree explanations were more expensive than Random Forest explanations; it is driven in part by the exceptionally small Decision Tree prediction denominator.

Random Forest incurred the largest explanation burden. Prediction times were 0.014148, 0.014128, 0.019230, 0.024609, and 0.024835~s across the five batches, while the reported TreeSHAP times were 0.125572, 1.293386, 13.077294, 132.182032, and 700.758801~s. The 5,000-sample TreeSHAP value was one recorded measurement. The corresponding overhead ratios were 8.88$\times$, 91.55$\times$, 680.05$\times$, 5,371.35$\times$, and 28,217.04$\times$.

XGBoost scaled more favorably. Prediction times were 0.000267, 0.000465, 0.000683, 0.001326, and 0.004109~s, while the reported TreeSHAP times were 0.003321, 0.009487, 0.034625, 0.277383, and 1.471088~s. The 5,000-sample TreeSHAP value was one recorded measurement. Its overhead ratios were 12.42$\times$, 20.41$\times$, 50.70$\times$, 209.15$\times$, and 357.99$\times$. Although explanation remained more expensive than prediction, both the absolute time and scaling behavior were substantially more favorable than for Random Forest.

Table~\ref{tab:cost-5000} summarizes the absolute and relative costs at the 5,000-sample operating point.

\begin{table}[htbp]
  \centering
  \caption{TreeSHAP computational cost at the 5,000-sample operating point.}
  \label{tab:cost-5000}
  \footnotesize
  \setlength{\tabcolsep}{4pt}
  \begin{tabular}{lrrrrr}
    \toprule
    Model & Prediction (s) & Explanation (s) & $\mu$s/sample & Samples/s & Overhead \\
    \midrule
    Decision Tree & 0.000433 & 5.642689 & 1,128.54 & 886.10 & 13,025.60$\times$ \\
    Random Forest & 0.024835 & 700.758801 & 140,151.76 & 7.135 & 28,217.04$\times$ \\
    XGBoost & 0.004109 & 1.471088 & 294.22 & 3,398.84 & 357.99$\times$ \\
    \bottomrule
  \end{tabular}
\end{table}

At 5,000 samples, Random Forest required approximately 476 times the absolute TreeSHAP time of XGBoost: 700.758801~s compared with 1.471088~s. Its per-sample explanation latency was 140,151.76~$\mu$s and its explanation throughput was 7.135 samples/s. Decision Tree required 1,128.54~$\mu$s per sample and produced 886.10 explanations/s, whereas XGBoost required 294.22~$\mu$s per sample and produced 3,398.84 explanations/s. These results establish that predictive efficiency does not imply explanation efficiency. Random Forest combined strong predictive performance with an exceptionally high TreeSHAP burden, while XGBoost combined strong predictive performance with substantially higher explanation throughput.

\subsection{TreeSHAP Numerical Validation}

Decision Tree and Random Forest additivity residuals were effectively at numerical precision. For XGBoost, an initial probability-space diagnostic was not a valid additivity test because the default TreeSHAP output is additive in the model's raw-margin, or log-odds, space. A separate validation used 50 observations, comprising 25 benign and 25 attack cases with 39 features. The raw-margin mean absolute error was $1.708441714072\times10^{-6}$ and the maximum absolute error was $4.333652668720\times10^{-6}$. After applying the sigmoid transformation, probability mean absolute error was $3.101912290149\times10^{-8}$ and maximum probability error was $2.140357095171\times10^{-7}$. Classification agreement was 1.000000. These results confirm numerical consistency of the TreeSHAP reconstruction; they do not establish causal validity or semantic correctness of the attributions.

The Stage-5A memory statistic was an observed before/after process resident-set-size difference. It was not a true peak-memory measurement and may not capture transient allocation peaks. Memory results are therefore not interpreted as peak requirements.
\section{Explanation Stability}

An explanation should not be treated as locally reliable solely because it can be generated. Prior work has shown that feature attributions can be sensitive to small input changes and adversarial manipulation \cite{ghorbani2019fragile,yeh2019infidelity,slack2020fooling}; IDS-specific work has likewise evaluated explanation behavior under adversarial perturbations \cite{arreche2024exai,munilla2026xairobustness}. The present protocol instead asks whether a small, semantically constrained perturbation that preserves both the predicted class and a similar confidence produces substantial changes in attribution structure. Stability was assessed for 100 base observations containing 50 benign and 50 attack cases. Ten candidate perturbations were generated for every base observation, with at most five accepted perturbations per sample and model.

For feature $j$, the perturbation scale was $\sigma_j=0.01\,\mathrm{IQR}_j$, where interquartile ranges were estimated from the Stage-3 validation data. Perturbations were restricted to nine continuous traffic statistics: \texttt{Rate}, \texttt{Tot sum}, \texttt{Min}, \texttt{Max}, \texttt{AVG}, \texttt{Std}, \texttt{Tot size}, \texttt{IAT}, and \texttt{Variance}. Binary protocol and flag features and discrete or count-like variables were not perturbed. Candidate values were clipped to validation-set 0.5th--99.5th percentile ranges and constrained to be non-negative. A candidate $x'$ was accepted only if
\begin{equation}
  f(x')=f(x)
  \quad\text{and}\quad
  |p(x')-p(x)|\leq 0.05,
\end{equation}
where $p(\cdot)$ is the model's attack confidence.

Four complementary metrics were used. Top-5 Jaccard similarity measures membership overlap among the five largest absolute attributions. Spearman correlation measures agreement between absolute-attribution rankings. Cosine similarity measures signed directional agreement between complete attribution vectors. Normalized $L_1$ change measures attribution-magnitude drift. Higher Jaccard, Spearman, and cosine values indicate greater stability, whereas lower normalized $L_1$ values indicate greater stability.

\subsection{Statistical Unit and Bootstrap Analysis}

Stages~6A and 6B generated multiple accepted perturbations from the same original base observation. Perturbation pairs sharing a base observation are correlated and therefore should not be treated as statistically independent units. The original pair-level summaries are retained below because they describe the distribution of accepted perturbation pairs, but Stage~6C uses the original base observation as the statistical unit for uncertainty estimation.

Stage~6C first averaged each stability metric across all accepted perturbations belonging to the same base observation. It then computed the mean across these per-base means and obtained percentile-bootstrap 95\% confidence intervals by resampling original base observations. The bootstrap used 10,000 repetitions and seed 2026. This analysis reused the recorded attribution comparisons and did not recompute SHAP values.

\subsection{Local Stability at Epsilon 0.01}

Table~\ref{tab:stability-main} reports the original perturbation-pair-level descriptive statistics. These summaries characterize a typical accepted perturbation pair but do not provide independent-sample uncertainty intervals.

\begin{table}[htbp]
  \centering
  \caption{Perturbation-pair descriptive explanation stability at $\epsilon=0.01$.}
  \label{tab:stability-main}
  \scriptsize
  \setlength{\tabcolsep}{3pt}
  \renewcommand{\arraystretch}{1.12}
  \begin{tabularx}{\textwidth}{>{\raggedright\arraybackslash}p{0.14\textwidth}*{7}{>{\centering\arraybackslash}X}}
    \toprule
    Model & Accepted pairs & Base samples & Median Jaccard & Mean Jaccard & Median Spearman & Median cosine & Median norm. $L_1$ \\
    \midrule
    Decision Tree & 480 & 100/100 & 1.000 & 0.871156 & 0.985613 & 0.999818 & 0.036678 \\
    Random Forest & 471 & 99/100 & 1.000 & 0.883834 & 0.994332 & 0.998687 & 0.044717 \\
    XGBoost & 463 & 98/100 & 0.666667 & 0.781960 & 0.967406 & 0.996619 & 0.109284 \\
    \bottomrule
  \end{tabularx}
\end{table}

Decision Tree and Random Forest exhibited very high local explanation stability. Both had a median Top-5 Jaccard similarity of 1.000, indicating complete median overlap among the five most highly attributed features. Random Forest produced the strongest attribution-rank stability, with a median Spearman correlation of 0.994332. Decision Tree had a median signed cosine similarity of 0.999818 and the smallest median normalized $L_1$ change, 0.036678.

XGBoost retained high global rank and directional consistency, with median Spearman and cosine values of 0.967406 and 0.996619. It was nevertheless comparatively more sensitive at the local top-$k$ and magnitude levels: median Top-5 Jaccard was 0.666667 and median normalized $L_1$ change was 0.109284. This does not support the categorical statement that XGBoost explanations were unstable. Rather, under the tested perturbations, their leading-feature membership and attribution magnitudes were more sensitive than those of Decision Tree and Random Forest.

\subsection{Base-Sample-Level Stability at Epsilon 0.010}

Table~\ref{tab:stability-base} reports the Stage-6C mean of the per-base means. Random Forest exhibited the strongest overall base-sample-level stability at $\epsilon=0.010$, combining the highest mean Top-5 Jaccard, Spearman, and signed-cosine values with the lowest mean normalized $L_1$ drift. Decision Tree remained highly stable in top-feature membership and attribution rank. XGBoost retained high rank and directional agreement, but showed lower Top-5 overlap and greater normalized attribution drift.

\begin{table}[htbp]
  \centering
  \caption{Base-sample-level explanation stability at $\epsilon=0.010$ with percentile-bootstrap 95\% confidence intervals. Each estimate is the mean of the per-base means.}
  \label{tab:stability-base}
  \scriptsize
  \setlength{\tabcolsep}{2.5pt}
  \renewcommand{\arraystretch}{1.28}
  \begin{tabularx}{\textwidth}{>{\raggedright\arraybackslash}p{0.13\textwidth}>{\centering\arraybackslash}p{0.035\textwidth}*{4}{>{\centering\arraybackslash}X}}
    \toprule
    Model & $n$ & Mean Top-5 Jaccard & Mean Spearman & Mean signed cosine & Mean normalized $L_1$ \\
    \midrule
    Decision Tree & 100 & \shortstack{0.866442\\{[0.845380, 0.886284]}} & \shortstack{0.981194\\{[0.978590, 0.983691]}} & \shortstack{0.953836\\{[0.939850, 0.966483]}} & \shortstack{0.162103\\{[0.129932, 0.196816]}} \\
    Random Forest & 99 & \shortstack{0.885426\\{[0.863299, 0.906718]}} & \shortstack{0.984711\\{[0.981771, 0.987482]}} & \shortstack{0.970922\\{[0.962716, 0.978569]}} & \shortstack{0.096899\\{[0.081574, 0.113380]}} \\
    XGBoost & 98 & \shortstack{0.783107\\{[0.762503, 0.803596]}} & \shortstack{0.960863\\{[0.956101, 0.965346]}} & \shortstack{0.954550\\{[0.945676, 0.962984]}} & \shortstack{0.189419\\{[0.169509, 0.210429]}} \\
    \bottomrule
  \end{tabularx}
\end{table}

The pair-level medians in Table~\ref{tab:stability-main} and the base-level means in Table~\ref{tab:stability-base} answer different questions. Pair-level medians describe the typical accepted perturbation pair, whereas base-level means give equal statistical weight to each original observation and expose observations with larger average drift. Their numerical differences are therefore not contradictory.

\subsection{Class-Conditioned Explanation Stability}

Table~\ref{tab:stability-class} stratifies the Stage-6C base-sample summaries by the original benign-versus-attack class. Across all three models, attack observations exhibited greater attribution-magnitude drift than benign observations. Attack observations also exhibited substantially lower signed-cosine similarity, particularly for Decision Tree and Random Forest. Rank and Top-5 behavior were model dependent, however, so these results do not support the claim that attack explanations were less stable on every metric.

\begin{table}[htbp]
  \centering
  \caption{Class-conditioned base-sample explanation stability at $\epsilon=0.010$. Bracketed values are percentile-bootstrap 95\% confidence intervals.}
  \label{tab:stability-class}
  \footnotesize
  \setlength{\tabcolsep}{5pt}
  \renewcommand{\arraystretch}{1.18}

  \textbf{Panel A---Feature-set and rank stability}\par\smallskip
  \begin{tabular}{llccc}
    \toprule
    Model & Class & $n$ & Mean Jaccard (95\% CI) & Mean Spearman (95\% CI) \\
    \midrule
    Decision Tree & Attack & 50 & \shortstack{0.867000\\{[0.846667, 0.888000]}} & \shortstack{0.982396\\{[0.979386, 0.985166]}} \\
                  & Benign & 50 & \shortstack{0.865885\\{[0.830378, 0.900282]}} & \shortstack{0.979993\\{[0.975611, 0.983950]}} \\
    \addlinespace
    Random Forest & Attack & 50 & \shortstack{0.852095\\{[0.825429, 0.878857]}} & \shortstack{0.976826\\{[0.972201, 0.981037]}} \\
                  & Benign & 49 & \shortstack{0.919436\\{[0.886199, 0.949433]}} & \shortstack{0.992757\\{[0.990812, 0.994481]}} \\
    \addlinespace
    XGBoost       & Attack & 50 & \shortstack{0.773651\\{[0.748633, 0.798605]}} & \shortstack{0.968088\\{[0.963037, 0.972425]}} \\
                  & Benign & 48 & \shortstack{0.792956\\{[0.759623, 0.827185]}} & \shortstack{0.953336\\{[0.945549, 0.960726]}} \\
    \bottomrule
  \end{tabular}

  \vspace{1em}
  \textbf{Panel B---Attribution-vector stability}\par\smallskip
  \begin{tabular}{llccc}
    \toprule
    Model & Class & $n$ & \shortstack{Mean signed cosine\\(95\% CI)} & \shortstack{Mean normalized $L_1$\\(95\% CI)} \\
    \midrule
    Decision Tree & Attack & 50 & \shortstack{0.908699\\{[0.888549, 0.927694]}} & \shortstack{0.278686\\{[0.233248, 0.326965]}} \\
                  & Benign & 50 & \shortstack{0.998973\\{[0.997981, 0.999560]}} & \shortstack{0.045520\\{[0.034810, 0.057681]}} \\
    \addlinespace
    Random Forest & Attack & 50 & \shortstack{0.945329\\{[0.932882, 0.956561]}} & \shortstack{0.144477\\{[0.121431, 0.169037]}} \\
                  & Benign & 49 & \shortstack{0.997038\\{[0.996163, 0.997851]}} & \shortstack{0.048350\\{[0.040989, 0.056242]}} \\
    \addlinespace
    XGBoost       & Attack & 50 & \shortstack{0.927818\\{[0.915846, 0.939559]}} & \shortstack{0.241949\\{[0.212975, 0.272373]}} \\
                  & Benign & 48 & \shortstack{0.982395\\{[0.975728, 0.988317]}} & \shortstack{0.134700\\{[0.115206, 0.154728]}} \\
    \bottomrule
  \end{tabular}
\end{table}

TP/FN/TN case-type diagnostics were also inspected, but the false-negative subgroups contained only three base observations per model and were therefore considered too small for inferential comparison.

\subsection{Sensitivity to Perturbation Magnitude}

The same protocol was repeated at $\epsilon\in\{0.005,0.010,0.020\}$. Table~\ref{tab:stability-epsilon} retains the original perturbation-pair descriptive summaries across these operating points.

\begin{table}[htbp]
  \centering
  \caption{Perturbation-pair descriptive explanation stability across perturbation strengths.}
  \label{tab:stability-epsilon}
  \small
  \setlength{\tabcolsep}{4pt}
  \begin{adjustbox}{max width=\textwidth}
  \begin{tabular}{lrrrrr}
    \toprule
    Model & $\epsilon$ & Mean Jaccard & Median Spearman & Median cosine & Median norm. $L_1$ \\
    \midrule
    Decision Tree & 0.005 & 0.912587 & 0.993313 & 0.999967 & 0.015123 \\
                  & 0.010 & 0.871156 & 0.985613 & 0.999818 & 0.036678 \\
                  & 0.020 & 0.827052 & 0.980547 & 0.999276 & 0.075716 \\
    \addlinespace
    Random Forest & 0.005 & 0.913073 & 0.997368 & 0.999519 & 0.025085 \\
                  & 0.010 & 0.883834 & 0.994332 & 0.998687 & 0.044717 \\
                  & 0.020 & 0.857246 & 0.989676 & 0.996896 & 0.066850 \\
    \addlinespace
    XGBoost       & 0.005 & 0.813585 & 0.974896 & 0.998229 & 0.076566 \\
                  & 0.010 & 0.781960 & 0.967406 & 0.996619 & 0.109284 \\
                  & 0.020 & 0.747146 & 0.957283 & 0.994686 & 0.137975 \\
    \bottomrule
  \end{tabular}
  \end{adjustbox}
\end{table}

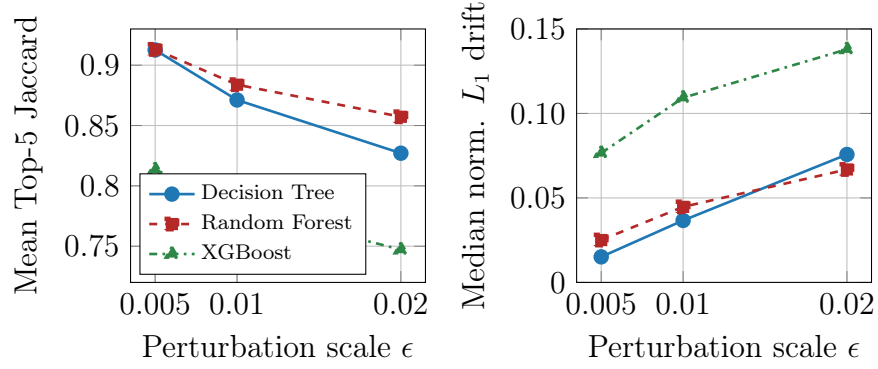
\begin{figure}[htbp]
  \centering
  \begin{tikzpicture}
    \begin{groupplot}[
      group style={group size=2 by 1,horizontal sep=2.0cm},
      width=0.40\textwidth,
      height=0.36\textwidth,
      xlabel={Perturbation scale $\epsilon$},
      xtick={0.005,0.010,0.020},
      scaled x ticks=false,
      xticklabel style={/pgf/number format/fixed,/pgf/number format/precision=3},
      grid=major,
      legend style={font=\scriptsize},
      legend cell align=left
    ]
      \nextgroupplot[
        ylabel={Mean Top-5 Jaccard},
        ymin=0.72,ymax=0.93,
        legend pos=south west
      ]
        \addplot[color=dtcolor,mark=*,mark size=2.4pt,line width=1pt,solid] coordinates {(0.005,0.912587) (0.010,0.871156) (0.020,0.827052)};
        \addlegendentry{Decision Tree}
        \addplot[color=rfcolor,mark=square*,mark size=2.3pt,line width=1pt,dashed] coordinates {(0.005,0.913073) (0.010,0.883834) (0.020,0.857246)};
        \addlegendentry{Random Forest}
        \addplot[color=xgbcolor,mark=triangle*,mark size=2.7pt,line width=1pt,dashdotted] coordinates {(0.005,0.813585) (0.010,0.781960) (0.020,0.747146)};
        \addlegendentry{XGBoost}
      \nextgroupplot[
        ylabel={Median norm. $L_1$ drift},
        ymin=0,ymax=0.15,
        ytick={0,0.05,0.10,0.15},
        yticklabels={0,0.05,0.10,0.15}
      ]
        \addplot[color=dtcolor,mark=*,mark size=2.4pt,line width=1pt,solid] coordinates {(0.005,0.015123) (0.010,0.036678) (0.020,0.075716)};
        \addplot[color=rfcolor,mark=square*,mark size=2.3pt,line width=1pt,dashed] coordinates {(0.005,0.025085) (0.010,0.044717) (0.020,0.066850)};
        \addplot[color=xgbcolor,mark=triangle*,mark size=2.7pt,line width=1pt,dashdotted] coordinates {(0.005,0.076566) (0.010,0.109284) (0.020,0.137975)};
    \end{groupplot}
  \end{tikzpicture}
  \caption{Perturbation-pair explanation stability as perturbation strength increases. The left panel reports mean Top-5 Jaccard similarity; the right panel reports median normalized $L_1$ attribution drift. Marker shapes and line styles distinguish models independently of color.}
  \label{fig:stability-epsilon}
\end{figure}

Figure~\ref{fig:stability-epsilon} visualizes the complementary perturbation-pair Top-5 overlap and normalized-magnitude responses across perturbation strengths.

At the perturbation-pair level, all models exhibited a monotonic degradation pattern as perturbation strength increased: mean Top-5 Jaccard, median Spearman, and median cosine similarity decreased, while normalized $L_1$ drift increased. Cosine similarity declined only moderately, showing that signed directional agreement can remain high even as top-feature membership and attribution magnitude change. XGBoost median Top-5 Jaccard remained 0.6667 at all tested levels because Top-5 Jaccard takes discrete values; the decreasing mean Jaccard provided greater sensitivity to distributional change.

The base-sample-level aggregation confirmed the same monotonic trend (Table~\ref{tab:stability-base-epsilon}). Mean Top-5 Jaccard decreased with perturbation strength for every model, while mean normalized $L_1$ drift increased.

\begin{table}[htbp]
  \centering
  \caption{Base-sample-level stability sensitivity across perturbation strengths. Values are means of the per-base means.}
  \label{tab:stability-base-epsilon}
  \small
  \setlength{\tabcolsep}{7pt}
  \begin{tabular}{lrrr}
    \toprule
    Model & $\epsilon$ & Mean Top-5 Jaccard & Mean normalized $L_1$ \\
    \midrule
    Decision Tree & 0.005 & 0.912762 & 0.119848 \\
                  & 0.010 & 0.866442 & 0.162103 \\
                  & 0.020 & 0.822479 & 0.203172 \\
    \addlinespace
    Random Forest & 0.005 & 0.912111 & 0.064236 \\
                  & 0.010 & 0.885426 & 0.096899 \\
                  & 0.020 & 0.858625 & 0.125522 \\
    \addlinespace
    XGBoost       & 0.005 & 0.817540 & 0.167672 \\
                  & 0.010 & 0.783107 & 0.189419 \\
                  & 0.020 & 0.748794 & 0.215619 \\
    \bottomrule
  \end{tabular}
\end{table}

Stages~6A and 6B reproduced the $\epsilon=0.010$ operating point exactly at the base-sample level. For Top-5 Jaccard, Spearman correlation, signed cosine similarity, and normalized $L_1$ change, the mean and maximum absolute differences between the two stages were 0.0 for all three models. This agreement supports deterministic internal reproducibility of the stability pipeline.
\section{Selective Explanation and Cost--Coverage Trade-offs}

Explain-all maximizes analytical coverage but may be unnecessary or computationally prohibitive. Selective prediction provides a broader methodological precedent for allocating automated actions according to confidence or risk \cite{geifman2017selective}; selective explanation instead applies this resource-allocation principle to XAI invocation while leaving the detector's class output unchanged. Stage~7A evaluated five policies. P0 explained every prediction. P1 explained predicted attacks only. P2 explained only uncertain predictions with $p(\mathrm{attack})\in[0.40,0.60]$. P3 explained predictions classified as attacks or falling in the uncertainty interval. P4 used a validation-calibrated risk trigger: every predicted attack was explained, together with predicted-benign cases whose attack probability exceeded a model-specific threshold.

P4 thresholds were derived from validation false negatives to target approximately 95\% false-negative explanation coverage. They were 0.004588 for Decision Tree, 0.010637 for Random Forest, and 0.006504 for XGBoost. Thresholds were calibrated without test labels.

\subsection{Selective Policies at the 95\% Target}

Table~\ref{tab:p4-balanced} reports the balanced-test results for the validation-calibrated P4 policy.

\begin{table}[htbp]
  \centering
  \caption{Validation-calibrated selective explanation on the balanced test at the approximately 95\% false-negative target.}
  \label{tab:p4-balanced}
  \footnotesize
  \setlength{\tabcolsep}{4pt}
  \begin{adjustbox}{max width=\textwidth}
  \begin{tabular}{lrrrrr}
    \toprule
    Model & Threshold & Explained (\%) & Saved (\%) & True-attack coverage (\%) & FN coverage (\%) \\
    \midrule
    Decision Tree & 0.004588 & 84.7639 & 15.2361 & 99.8942 & 96.1978 \\
    Random Forest & 0.010637 & 77.4214 & 22.5786 & 99.8483 & 95.0538 \\
    XGBoost & 0.006504 & 76.5269 & 23.4731 & 99.8599 & 94.9459 \\
    \bottomrule
  \end{tabular}
  \end{adjustbox}
\end{table}

On the balanced test, P1 reduced explanation workload by approximately 51\%, but its false-negative explanation coverage was exactly zero because every missed attack was predicted benign. P2 saved more than 99.5\% of workload but explained almost none of the true attacks, making it unsuitable as a standalone security policy. P3 recovered only a small fraction of false negatives, indicating that many missed attacks were not simply close to the 0.5 decision boundary. These outcomes motivate a risk-calibrated trigger rather than a narrow uncertainty-only policy.

P4 retained nearly complete true-attack coverage while reducing workload. XGBoost produced the largest balanced-test saving at the 95\% target, 23.4731\%, followed by Random Forest at 22.5786\% and Decision Tree at 15.2361\%. The achieved false-negative coverage varied slightly around the calibration target because thresholds were derived from validation data rather than tuned on test labels.

\subsection{Prevalence Dependence}

Selective-explanation savings were substantially smaller on the natural attack-dominated test distribution. At the P4 95\% target, Decision Tree saved 1.7169\% of explanation workload while covering 99.8838\% of true attacks and 95.8812\% of false negatives. Random Forest saved 2.5255\%, with 99.8480\% true-attack coverage and 95.1367\% false-negative coverage. XGBoost saved 2.6071\%, with 99.8624\% true-attack coverage and 95.1115\% false-negative coverage.

Selective-explanation savings are therefore prevalence-dependent. When attacks dominate the workload, a security policy that explains almost every attack necessarily retains a large explanation burden. The smaller natural-test saving is an operational property of the workload and policy objective rather than a failure of selective explanation.

\subsection{Cost--Coverage Frontier}

Stage~7B swept target validation false-negative explanation coverage over 50\%, 70\%, 80\%, 90\%, 95\%, and 99\%. For target $C$, the trigger threshold was the $(1-C)$-quantile of validation false-negative attack probabilities. Test labels were not used for threshold calibration.

\begin{figure}[htbp]
  \centering
  \begin{tikzpicture}
    \begin{axis}[
      width=0.90\textwidth,
      height=0.52\textwidth,
      xlabel={False-negative explanation coverage (\%)},
      ylabel={Compute saved (\%)},
      xmin=47,xmax=100,
      ymin=0,ymax=50,
      grid=major,
      legend pos=south west,
      legend cell align=left,
      legend style={font=\small}
    ]
      \addplot[color=dtcolor,mark=*,mark size=2.4pt,line width=1pt,solid] coordinates {
        (49.7802,46.092971) (70.0659,40.905922) (83.2527,34.579514)
        (90.1538,28.351263) (96.1978,15.236097) (99.0330,5.248207)};
      \addlegendentry{Decision Tree}
      \addplot[color=rfcolor,mark=square*,mark size=2.3pt,line width=1pt,dashed] coordinates {
        (49.0227,47.383083) (68.8273,42.811581) (79.3777,38.481344)
        (90.2074,31.174891) (95.0538,22.578633) (99.1623,6.714145)};
      \addlegendentry{Random Forest}
      \addplot[color=xgbcolor,mark=triangle*,mark size=2.7pt,line width=1pt,dashdotted] coordinates {
        (49.3048,47.477876) (69.2783,43.478806) (79.5851,39.653728)
        (89.8698,32.217316) (94.9459,23.473057) (99.0951,8.175497)};
      \addlegendentry{XGBoost}
    \end{axis}
  \end{tikzpicture}
  \caption{Balanced-test cost--coverage frontier using validation-calibrated false-negative risk thresholds. Marker shapes and line styles distinguish models independently of color.}
  \label{fig:cost-coverage}
\end{figure}
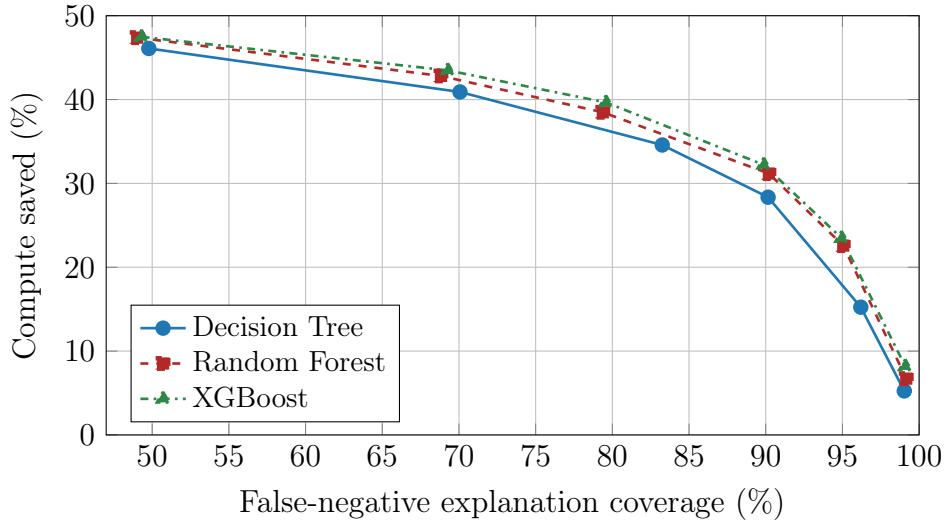

Figure~\ref{fig:cost-coverage} reports the balanced-test frontier produced by these validation-calibrated thresholds.

Table~\ref{tab:frontier-key} summarizes the principal 90\% and 95\% operating points.

\begin{table}[htbp]
  \centering
  \caption{Balanced-test cost--coverage frontier at the 90\% and 95\% validation targets.}
  \label{tab:frontier-key}
  \small
  \begin{adjustbox}{max width=\textwidth}
  \begin{tabular}{lrrrr}
    \toprule
    & \multicolumn{2}{c}{90\% target} & \multicolumn{2}{c}{95\% target} \\
    \cmidrule(lr){2-3}\cmidrule(lr){4-5}
    Model & Test FN coverage (\%) & Saved (\%) & Test FN coverage (\%) & Saved (\%) \\
    \midrule
    Decision Tree & 90.1538 & 28.351263 & 96.1978 & 15.236097 \\
    Random Forest & 90.2074 & 31.174891 & 95.0538 & 22.578633 \\
    XGBoost & 89.8698 & 32.217316 & 94.9459 & 23.473057 \\
    \bottomrule
  \end{tabular}
  \end{adjustbox}
\end{table}

At approximately 90\% false-negative coverage on the balanced test, compute savings were 28.35\% for Decision Tree, 31.17\% for Random Forest, and 32.22\% for XGBoost. At approximately 95\% coverage, the corresponding savings were 15.24\%, 22.58\%, and 23.47\%. All tested frontier points were non-dominated under the paired objectives of maximizing false-negative explanation coverage and compute saving. No single threshold is therefore universally optimal; the operating point depends on deployment risk tolerance and resource budget.

The natural-test frontier was compressed by attack prevalence. At the 90\% target, Decision Tree, Random Forest, and XGBoost saved 3.25\%, 3.58\%, and 3.66\%, respectively. At the 95\% target, savings were 1.72\%, 2.53\%, and 2.61\%. The lower savings primarily reflect the requirement to explain nearly all attacks in a workload dominated by attacks.

Stage-7 full-test explanation times are throughput-based projections from the Stage-5A 5,000-sample operating point, not measured full-test runtimes. Under an explain-all host-runtime projection, Random Forest required approximately 120.30~h for the natural test, compared with approximately 0.253~h for XGBoost. These projected workloads emphasize architecture-dependent XAI cost but do not represent direct measurements on IoT hardware.
\section{Discussion}

\subsection{Detection Accuracy Is Not an Adequate Deployment Criterion}

The predictive experiments show why accuracy cannot determine deployment suitability in isolation. XGBoost, Random Forest, and Decision Tree produced closely grouped F1 scores, yet differed in false-positive behavior, serialized size, host training time, explanation cost, and local stability. Logistic Regression further showed that a threshold optimized for F1 under an attack-dominated distribution can create an operationally undesirable false-positive rate. Model selection must therefore begin with predictive quality but cannot end there.

\subsection{Explanation Cost Is Architecture-Dependent}

The cost results expose a model-architecture effect that is not visible in detection metrics. At 5,000 samples, Random Forest TreeSHAP required 700.758801~s, whereas XGBoost required 1.471088~s. Random Forest explanations were highly stable, but their absolute cost and throughput were substantially less favorable. XGBoost explanations were dramatically cheaper and scaled more effectively. Decision Tree illustrates why overhead ratios cannot be interpreted alone: its 13,025.60$\times$ ratio was produced by a very small prediction time, while its absolute TreeSHAP time remained far below that of Random Forest.

\subsection{Stability Is Multi-Dimensional}

Stability depends on which aspect of an explanation is examined. Top-5 Jaccard measures feature-set membership, Spearman correlation measures rank agreement, signed cosine measures directional alignment, and normalized $L_1$ captures magnitude drift. XGBoost demonstrates why a single measure is insufficient: cosine similarity remained 0.996619 at $\epsilon=0.01$, even though median Top-5 Jaccard was 0.666667 and normalized $L_1$ drift was 0.109284. High directional agreement can therefore coexist with greater turnover among the most prominent features and larger changes in attribution magnitude.

Base-sample aggregation prevents observations with more accepted perturbations from receiving greater statistical weight and identifies Random Forest as the strongest overall base-level stability profile at $\epsilon=0.010$. The class-conditioned analysis adds a separate qualification: attack observations had greater magnitude drift and lower signed-cosine agreement across all three models, but Top-5 membership and rank differences varied by architecture. Stability comparisons must therefore specify the metric, statistical unit, perturbation strength, and traffic class.

\subsection{Selective Explanation as Resource Allocation}

Selective explanation reframes XAI as a computational security resource rather than an automatic post-processing step. Explain-all provides maximum coverage at maximum cost. Explaining predicted attacks alone yields substantial savings on a balanced workload but never explains false negatives. Uncertainty-only invocation is inexpensive but provides inadequate attack coverage, while an attack-or-uncertain policy recovers only a limited fraction of missed attacks. Validation-calibrated risk creates a tunable compromise between false-negative explanation coverage and computational load without using test labels for threshold selection.

\subsection{Practical Deployment Implications}

The host-based measurements suggest three operational profiles that warrant subsequent device-level evaluation. For a lightweight gateway, Decision Tree may be attractive because its serialized model is compact and its absolute TreeSHAP cost is low relative to Random Forest, while its predictive performance remains competitive. This profile is a hypothesis for constrained-hardware evaluation, not evidence that the reported host timings transfer directly to an IoT gateway.

In a false-alarm-sensitive environment, Random Forest offers the lowest observed false-positive rate, but its explanation throughput makes an explain-all policy difficult to justify at high event volumes. Selective invocation is particularly important for this architecture because each explained case consumes substantially more host time. For high-throughput explainable monitoring, XGBoost offers the strongest combined predictive and explanation-cost profile on the tested host. The appropriate profile still depends on the deployed traffic prevalence, explanation budget, acceptable false-negative coverage, and the performance of the target hardware.

\subsection{Model-Level Trade-off}

Table~\ref{tab:model-synthesis} synthesizes the four evaluated dimensions: predictive performance, explanation cost, stability, and selective explanation.

\begin{table}[htbp]
  \centering
  \caption{Cross-dimensional synthesis of the evaluated models.}
  \label{tab:model-synthesis}
  \footnotesize
  \renewcommand{\arraystretch}{1.15}
  \begin{tabularx}{\textwidth}{>{\raggedright\arraybackslash}p{0.13\textwidth}*{4}{>{\raggedright\arraybackslash}X}}
    \toprule
    Model & Predictive profile & Explanation cost & Stability & Selective-XAI implications \\
    \midrule
    Decision Tree & Competitive detection with a compact serialized model & Low absolute TreeSHAP cost, despite a large ratio to very fast prediction & Very high feature-set, rank, directional, and magnitude stability & Compact baseline for further resource-constrained hardware evaluation \\
    Random Forest & Strong detection and the lowest false-positive rate among the tree models & Extremely high TreeSHAP cost and lowest explanation throughput & Highest rank stability and high top-feature overlap & Selective invocation is especially important because each explanation is costly \\
    XGBoost & Strongest overall predictive profile within the evaluated task & Lowest TreeSHAP cost at scale and highest explanation throughput & High rank and directional consistency, with greater local top-$k$ and magnitude sensitivity & Most favorable combined predictive and explanation-cost profile on the evaluated host \\
    \bottomrule
  \end{tabularx}
\end{table}

Within the evaluated host-based configuration and CICIoT2023 binary task, XGBoost presented the most favorable combined predictive/explanation-cost profile. This is not a universal model ranking. Random Forest offered the lowest false-positive rate and strongest rank stability, while Decision Tree remained competitive with low absolute explanation cost and compact serialized size. The appropriate choice depends on the relative cost assigned to missed attacks, false alarms, explanatory throughput, local attribution sensitivity, and available resources.

\begin{figure}[htbp]
  \centering
  \begin{tikzpicture}[
    node distance=8mm and 7mm,
    box/.style={draw,rounded corners,align=center,minimum width=3.2cm,minimum height=0.85cm,fill=gray!8,font=\small},
    arrow/.style={-{Latex[length=2.5mm]},line width=0.8pt}
  ]
    \node[box] (detect) {Detection\\Performance};
    \node[box,right=of detect] (predict) {Prediction\\Cost};
    \node[box,right=of predict] (explain) {Explanation\\Cost};
    \node[box,below=of explain] (stable) {Explanation\\Stability};
    \node[box,left=of stable] (select) {Selective Explanation\\Utility};
    \node[box,left=of select,fill=dtcolor!10] (operational) {Operational\\Utility};
    \draw[arrow] (detect) -- (predict);
    \draw[arrow] (predict) -- (explain);
    \draw[arrow] (explain) -- (stable);
    \draw[arrow] (stable) -- (select);
    \draw[arrow] (select) -- (operational);
  \end{tikzpicture}
  \caption{Overall operational trade-off: predictive performance is the first stage, rather than the complete criterion, in evaluating resource-aware explainable intrusion detection.}
  \label{fig:operational-tradeoff}
\end{figure}
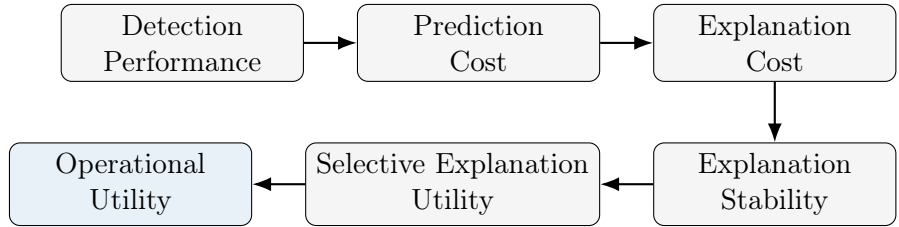

Figure~\ref{fig:operational-tradeoff} summarizes the progression from detection performance to operational utility.

\section{Limitations and Threats to Validity}

\textbf{Dataset scope.} The primary empirical conclusions are based on CICIoT2023 and binary attack-versus-benign classification. The retained original labels and attack families support diagnostics, but the reported models were not evaluated as multiclass family classifiers.

\textbf{External validity.} No external-dataset validation has yet been conducted. The results do not establish cross-dataset generalization to Edge-IIoTset or another traffic environment.

\textbf{Host-based timing.} Training, prediction, and TreeSHAP measurements were obtained on the experimental host, not on a Raspberry Pi, production IoT gateway, or other constrained device. The measurements therefore do not demonstrate direct edge-device feasibility and are hardware dependent.

\textbf{Memory measurement.} Stage-5 memory observations used before/after process resident-set-size differences. This method can miss transient allocation peaks and must not be interpreted as true peak-memory measurement.

\textbf{TreeSHAP scope.} Explanation-cost conclusions concern TreeSHAP for the evaluated Decision Tree, Random Forest, and XGBoost models. They should not be generalized automatically to LIME, model-agnostic SHAP variants, counterfactual methods, or other XAI procedures.

\textbf{Perturbation validity.} Stage-6 perturbations were restricted to selected continuous traffic statistics and bounded by validation-derived percentiles. They provide a controlled local sensitivity test but do not constitute complete protocol-level or causal traffic generation.

\textbf{Interpretation of stability.} Stable SHAP attributions do not prove that model reasoning is causal, semantically correct, or operationally appropriate. Stability measures repeatability under the tested neighborhood and explainer configuration.

\textbf{Statistical uncertainty.} Predictive metrics are point estimates from fixed fitted models and deterministic test partitions; repeated model training and formal significance testing were not performed. Accordingly, small differences among predictive models are interpreted descriptively rather than as evidence of statistical superiority. For the Stage-6 stability analysis, dependence among multiple perturbations of the same observation was addressed through base-sample aggregation and 10,000-fold bootstrap confidence intervals across original base observations.

\textbf{Projected selective-explanation cost.} Full-dataset Stage-7 explanation times are throughput-based estimates derived from Stage-5A measurements at the 5,000-sample operating point. They are host-runtime projections rather than directly measured full-test runtimes.

\textbf{Prevalence dependence.} Selective-policy utility changed substantially between the balanced and natural test distributions. Savings measured under one prevalence should not be transferred to another workload without recalculation.

\textbf{Operational utility.} Utility in this study refers to computational workload, security-case coverage, and cost--coverage trade-offs under selective explanation. No human-subject study was performed, and the results do not establish improvements in analyst trust, decision quality, or response effectiveness.

\textbf{Binary focus.} Multiclass attack-family detection may produce different predictive errors, explanation costs, local stability patterns, and selection-policy trade-offs.

\section{Conclusion}

This study evaluated explainable IoT intrusion detection across predictive effectiveness, explanation cost, explanation stability, and selective invocation. For RQ1, the tree-based models achieved strong binary detection performance on the leakage-safe CICIoT2023 corpus. XGBoost produced the strongest overall predictive profile under the reported metrics, while Random Forest produced particularly low false-positive rates. These results also confirmed that accuracy alone obscures operationally important differences in recall and benign false alarms.

For RQ2, TreeSHAP cost differed by orders of magnitude across model architectures. At 5,000 samples, Random Forest required approximately 700.8~s, compared with approximately 5.64~s for Decision Tree and 1.47~s for XGBoost. Prediction speed alone did not determine explanation speed, and relative overhead ratios required interpretation alongside absolute latency and throughput.

For RQ3, explanations were generally stable under the tested prediction-preserving perturbations, but stability was model dependent and multi-dimensional. Base-sample aggregation identified Random Forest as the strongest overall stability profile at $\epsilon=0.010$, while Decision Tree also exhibited strong top-feature and rank stability. XGBoost maintained high directional and rank consistency while showing greater top-$k$ membership turnover and magnitude sensitivity. Stability declined systematically as perturbation strength increased. Attack observations exhibited greater magnitude drift and lower signed-cosine agreement than benign observations across all three models, although rank and Top-5 differences remained architecture dependent.

For RQ4, validation-calibrated selective explanation reduced workload while retaining security-relevant coverage. On the balanced test, approximately 90\% false-negative explanation coverage permitted savings of roughly 28--32\%, depending on model. At approximately 95\% coverage, savings ranged from roughly 15--23\%. Benefits were much smaller on the attack-heavy natural test because a policy that explains nearly every attack must retain most of the workload.

The operational value of explainable IoT intrusion detection cannot be determined from detection accuracy or explanation availability alone. Predictive quality, architecture-dependent explanation cost, local stability, workload prevalence, and security-relevant coverage must be evaluated jointly. The present results establish these trade-offs for the evaluated host configuration and CICIoT2023 binary task without claiming universal model superiority or direct constrained-device feasibility.

\section*{Funding}

This research did not receive any specific grant from funding agencies in the public, commercial, or not-for-profit sectors.

\section*{CRediT authorship contribution statement}

Abdurrahman Tolay: Conceptualization, Methodology, Software, Validation, Formal analysis, Investigation, Data curation, Visualization, Writing~-- original draft, Writing~-- review \& editing.

\section*{Declaration of competing interest}

The author declares no known competing financial interests or personal relationships that could have appeared to influence the work reported in this paper.

\section*{Data availability}

The CICIoT2023 dataset used in this study is publicly available from its original source and is cited in the manuscript. The preprocessing, corpus-audit, predictive-evaluation, explanation-cost, stability-analysis, and selective-explanation scripts, together with trained model artifacts and derived experimental outputs, are publicly available through the reproducibility repository at GitHub (https://github.com/AbdurrahmanTolay/\allowbreak resource-aware-iot-xai) and its archived Zenodo release (DOI: 10.5281/zenodo.21879038).

\section*{Declaration of generative AI and AI-assisted technologies in the manuscript preparation process}

During the preparation of this manuscript, the author used Prism for language checking, editorial refinement, and organization of the LaTeX manuscript. The research design, experimental methodology, implementation, data processing, analyses, numerical results, scientific interpretation, and final content were developed, verified, and approved by the author, who takes full responsibility for the work.

\bibliographystyle{elsarticle-num}
\bibliography{references}

\end{document}